\documentclass[aps,prb,twocolumn,superscriptaddress,showpacs]{revtex4-2}
\usepackage{graphicx}
\usepackage{mathrsfs}
\usepackage{bm}
\usepackage{amsfonts}
\usepackage{amsmath}
\usepackage{color}
\usepackage{dcolumn}
\usepackage{epstopdf}
\usepackage{dsfont}
\usepackage{amssymb}
\usepackage{tabularx}
\usepackage{array}
\usepackage{float}
\usepackage{colordvi}
\usepackage{verbatim}
\usepackage{hyperref}
\hypersetup{
	colorlinks=true,
	linkcolor=blue,
	filecolor=blue,
	urlcolor=blue,
	citecolor=blue,
}
\usepackage{amsfonts}
\usepackage{multirow}
\usepackage{bigstrut}
\usepackage{rotating}
\usepackage{booktabs}
\usepackage{braket}
\usepackage{enumerate}

\begin{document}
\title{Two-dimensional Dirac semimetals with tunable edge states}

\author{Lizhou Liu}
\affiliation{College of Physics, Hebei Normal University, Shijiazhuang 050024, China}

\author{Cheng-Ming Miao}
\affiliation{International Center for Quantum Materials, School of Physics, Peking University, Beijing 100871, China}

\author{Qing-Feng Sun}
\affiliation{International Center for Quantum Materials, School of Physics, Peking University, Beijing 100871, China}
\affiliation{Hefei National Laboratory, Hefei 230088, China}

\author{Ying-Tao Zhang}
\email[Correspondence author:~~]{zhangyt@mail.hebtu.edu.cn}
\affiliation{College of Physics, Hebei Normal University, Shijiazhuang 050024, China}

\date{\today}

\begin{abstract}
  We theoretically propose a design for two-dimensional Dirac semimetals using a bilayer-modified Bernevig-Hughes-Zhang (BHZ) model. By introducing new sites into the BHZ model, we engineer flat bands at the Fermi energy. In the bilayer system, interlayer coupling separates these flat bands, resulting in two Dirac points that preserve time-reversal and inversion symmetries. Two Dirac points are connected by a one-dimensional Fermi arc edge state, whose bound nature is confirmed by quantized transmission resonance peaks. Notably, the position of the Dirac points can be precisely tuned by adjusting interlayer coupling strengths and symmetries.
\end{abstract}

\maketitle
\section{Introduction}
Since the groundbreaking discovery of graphene in 2004~\cite{Novoselov2004}, the study of two-dimensional semimetals has rapidly expanded, driven by their unique electronic properties. These materials, particularly topological semimetals, exhibit a range of intriguing phenomena stemming from symmetry-protected band crossings at the Fermi energy. These phenomena include ultrahigh electron mobility~\cite{Shekhar2015, Liang2015}, negative magnetoresistance~\cite{Huang2015, Wang2016, Arnold2016}, colossal photovoltaic responses~\cite{Osterhoudt2019, Ma2019}, and chiral anomalies~\cite{Hirschberger2016, Son2013}.
Topological semimetals are broadly categorized into nodal semimetals and nodal line semimetals~\cite{Du2017, Kim2015, Yang2018, Fang2016}, depending on the dimensionality of their band crossings. Within nodal semimetals, Dirac semimetals feature fourfold degenerate points due to the preservation of time-reversal and inversion symmetries~\cite{Burkov2017, Weng2016, Wang2012, Wang2013}. In contrast, Weyl semimetals are characterized by pairs of twofold degenerate Weyl points, arising from broken symmetries~\cite{Wan2011, Xu2011, Chen2015}. However, spin-orbit coupling in two-dimensional semimetals often disrupts these band crossings, leading to the formation of topological insulator phases~\cite{Kane2005, Kane2005a, Bernevig2006}.

The search for two-dimensional semimetals is complicated by the prevalence of heavy elements in most candidate materials, which induce strong spin-orbit coupling. This challenge has prompted a shift in research focus towards three-dimensional systems.
Within these systems, it has been demonstrated that a Dirac semimetal phase can emerge at the boundary between a topological phase and a trivial insulator phase, provided both time-reversal and inversion symmetries are maintained~\cite{Murakami2007}.
When certain perturbations break either time-reversal symmetry~\cite{Wan2011, Xu2011, Xu2018} or inversion symmetry~\cite{Lv2015, Xu2015, Huang2015a, Weng2015}, the Dirac node can split into two Weyl nodes with opposite chirality. This splitting gives rise to the Weyl semimetal phase, which is distinguished by phenomena such as the appearance of Fermi arcs on the surface.

To further explore the distinctive properties of two-dimensional semimetals, it is crucial to investigate those that remain robust against spin-orbit coupling. The pioneering work by Young and Kane demonstrated that nonsymmorphic symmetry can stabilize two-dimensional Dirac points at the Brillouin zone boundary even in the presence of spin-orbit coupling~\cite{Young2015}. Following this, numerous two-dimensional semimetals have been predicted, including Dirac semimetals~\cite{Guan2017, Li2018, Jin2020, Young2017, Chen2017, Li2019}, Weyl semimetals~\cite{Matveeva2019}, and nodal line semimetals~\cite{Cao2023, Wu2022}.
Nonsymmorphic operators, with their higher-dimensional projective representations, lead to the formation of two-dimensional Weyl nodes characterized by a local Chern number or quantized Berry phase. These nodes are associated with finite edge states that terminate at the projections of two Weyl points in the one-dimensional Brillouin zone~\cite{Matveeva2019, Young2017}.
Moreover, nontrivial edge states are known to exhibit unique electronic transport properties, such as robust quantized conductance and spin-polarized currents~\cite{Hasan2010, Qi2011, Peres2010, Yang2011, Burkov2011}.
However, in two-dimensional Dirac semimetals, the Berry curvature vanishes throughout the entire Brillouin zone when both inversion and time-reversal symmetries are present~\cite{Young2015, Xiao2010}, resulting in the absence of edge states.
On the other hand, Dirac points in two-dimensional semimetals are mainly pinned at high symmetry points.
Fixed positions limit the functionality of Dirac points, for example, Dirac points in graphene are pinned to K and K' points.
Therefore, it is necessary to design two-dimensional Dirac semimetals with spin-orbit coupling, tunable edge states, and tunable positions of Dirac points.

In this paper, we propose a method to construct Dirac semimetals with edge states using a bilayer-modified Bernevig-Hughes-Zhang (BHZ) model. The single-layer model, depicted in Fig.~\ref{fig1}(a), features newly introduced sites (highlighted in blue) that are deposited into the vacancies of the BHZ model, with a nearest-neighbor hopping amplitude denoted as \( t_2 \). By coupling the top and bottom layers of the modified BHZ models in various configurations, we demonstrate that Dirac points emerge in the two-dimensional bulk, protected by mirror symmetry \( M_y \).
The two Dirac points in the Brillouin zone are symmetric about the $\Gamma$ point and the distance between them can be adjusted by the interlayer coupling strengths.
The symmetry properties of bilayer model, such as \( M_x \) and \( M_{xy} \), can be manipulated by adjusting the parameters, allowing precise control over the directions of Dirac points.
Additionally, we find that a flat edge state connects these Dirac points, resulting in anisotropic edge states. The presence of bound edge states is confirmed by quantized transmission resonance peaks.
Finally, we show that magnetic deposition will cause each Dirac point in the bilayer system to cleave into two Weyl points.

The remaining of the paper is organized as follows.
In Sec.~\ref{SecII}, we introduce the tight-binding Hamiltonian of the BHZ model with deposited atoms.
In Sec.~\ref{SecIII}, we study the bulk band and energy band structure of Dirac semimetals in bilayer systems.
In Sec.~\ref{SecIV}, we show the Weyl points in the bilayer modified BHZ model and illustrate the existence of edge states.
Finally, a conclusion is presented in Sec.~\ref{SecV}.

\section{model and hamiltonian } \label{SecII}

\begin{figure}
  \centering
  \includegraphics[width=8cm,angle=0]{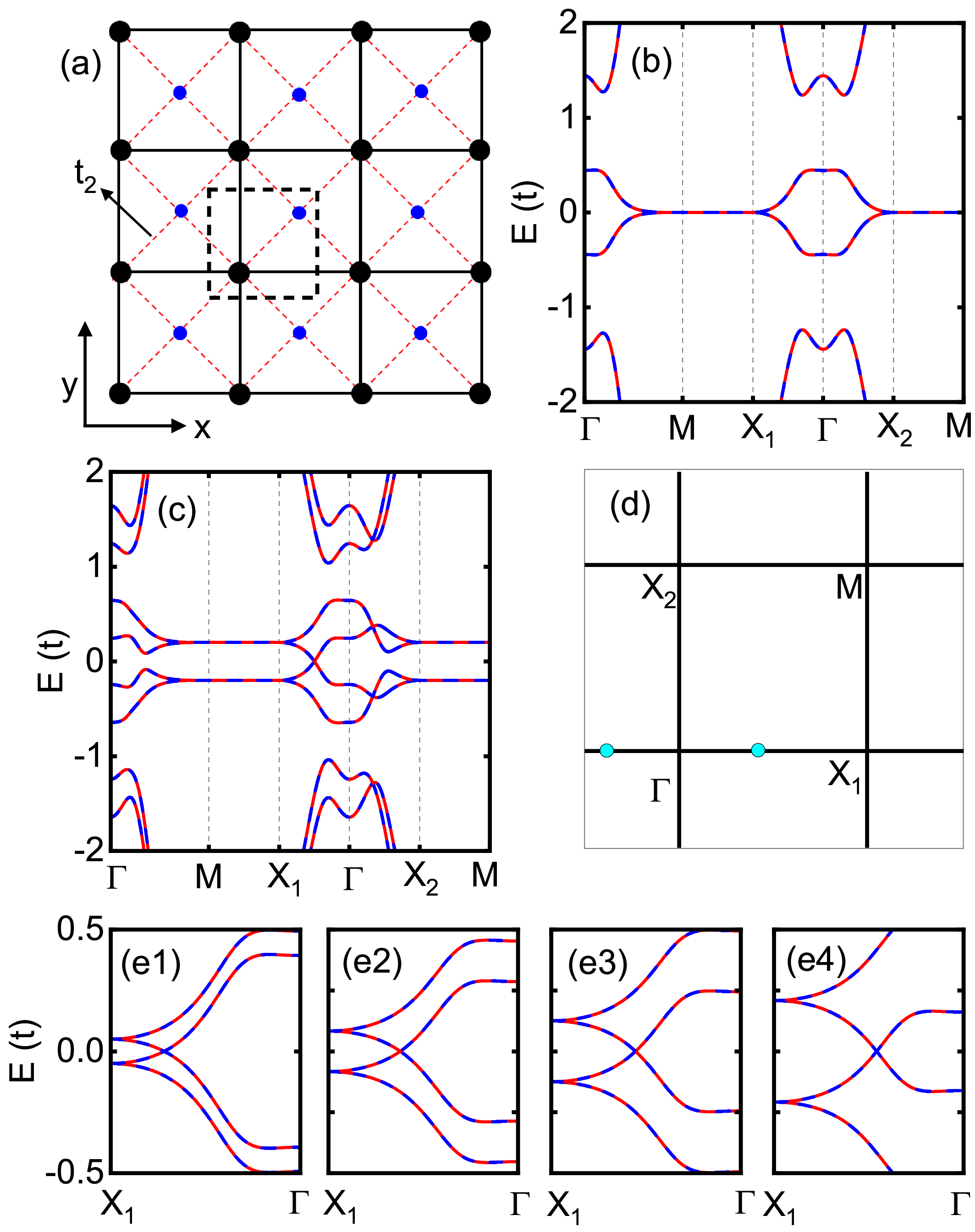}
  \caption{(a) The two-dimensional lattice consists of two types of sites, depicted by black and blue spheres, respectively. The square grid formed by the black spheres represents the BHZ model, while the blue spheres denote deposited correction sites located within the vacancies. The dashed rectangular area outlines the unit cell. The coupling amplitude of the deposited site to its nearest neighbor in the BHZ model is denoted by $t_2$.
  (b) The bulk energy spectrum of the single-layer model along the high-symmetry points with $t = 1, \epsilon=-1, \lambda_x=\lambda_y=1, t_2=0.2$.
  (c) The bulk energy spectrum of the bilayer model along the high-symmetry points with $\eta=0.2$.
  (d) Dirac points in the Brillouin Zone are marked in cyan.
 In the presence of the time-reversal and inversion symmetries, each line is twofold degenerate, the red solid lines indicate the spin-up bands and the blue dashed lines indicate the spin-down bands.
  High-symmetry points $\Gamma=(0,0), M=(\pi,\pi), X_1=(\pi,0), X_2=(0,\pi)$.
 (e1)-(e4) The bulk band structure of the bilayer model along the high symmetric path $X_1-\Gamma$ with different interlayer coupling strengths $(\eta=0.05, 0.1, 0.15, 0.25)$.}
  \label{fig1}
\end{figure}

The bilayer-modified BHZ model can be described by the following tight-binding Hamiltonian:
\begin{align}
H(\mathbf {k})=
\left( \begin{matrix}
H^T &  \eta \sigma_0 s_0 \\
\eta \sigma_0 s_0^* &  H^B \\
\end{matrix} \right),
\label{eq1}
\end{align}
where $H^T$ and $H^B$ are the momentum space Hamiltonians of the top and bottom modified BHZ models as shown in Fig.~\ref{fig1}(a), with a coupling intensity $\eta$ between them.
The $H^T$ and $H^B$ are given by:
\begin{align}
H^{T,B}=
\left( \begin{matrix}
H_{BHZ}^{T,B} &  4 t_2 \cos \frac{k_x}{2} \cos \frac{k_y}{2} \sigma_z s_0 \\
4 t_2 \cos \frac{k_x}{2} \cos \frac{k_y}{2} \sigma_z s_0  &  B_x s_x\sigma_z + \mu \sigma_0 s_0 \\
\end{matrix} \right),
\label{eq2}
\end{align}
where $\sigma_i$ and $s_i$ for $i \in \left \{x, y, z\right\}$ are Pauli matrices acting in orbital and spin space, respectively.
$H_{BHZ}^T$ and $H_{BHZ}^B$ are the momentum space Hamiltonians of the BHZ models.
$t_2$ is the nearest-neighbor coupling term between the deposited site and the site in the BHZ model,
$B_{x}$ and $\mu$ denote the magnetic and chemical potentials of the deposited sites, respectively. $B_{x/z} = \mu = 0$ unless otherwise stated.
In momentum space, the Hamiltonian of a single BHZ models layer can then be written as $H_0=\sum_k \Psi_k^{\dag} H_0(k) \Psi_k$ in the basis $\Psi_k = (\psi_{k11}, \psi_{k1\bar{1}}, \psi_{k\bar{1}1}, \psi_{k\bar{1}\bar{1}} )$, where $\psi_{k \sigma s} (\psi_{k \sigma s}^{\dag})$ destroys (creates) an electron with in-plane momentum $k=(k_x, k_y)$, orbital degree of freedom $\sigma \in \left \{ 1, \bar{1} \right \}$, and spin $s \in \left \{ 1, \bar{1} \right \}$.
The top layer BHZ Hamiltonian is given by ~\cite{Bernevig2006}:
\begin{align}
H_{BHZ}^T&=(4t+ \epsilon - 2t \cos k_x - 2t \cos k_y ) \sigma_z \nonumber \\
& + \lambda_x \sin k_x \sigma_x s_z + \lambda_y \sin k_y \sigma_y,
\label{eq3}
\end{align}
where $t$ is the nearest-neighbor intraorbital hopping,
$\epsilon$ describes a relative energy shift between the particle-like $(\sigma = 1)$ and holelike $(\sigma = \bar{1})$ bands,
and $\lambda_{x,y}$ measure the kinetic energy.
To satisfy different mirror symmetries, in the following different subsections, we select different Hamiltonians $H_{BHZ}^B$ and discuss them separately.
We demonstrate that if $B_{x}=0$ the Hamiltonian $H(\textbf{k})$ in Eq.(\ref{eq1}) maintains time-reversal symmetry $\mathcal{T} = i s_y \mathcal{K}$ ($\mathcal{K}$ is the complex conjugation) and inversion symmetries $\mathcal{I}\sigma_z$, since $(\mathcal{T} \mathcal{I})^2=-1$, the bulk bands are Kramers degenerate for all $\mathbf{k}$.

\section{two-dimensional Dirac Semimetals in bilayer systems}\label{SecIII}

It is well known that the BHZ model without spin-orbit coupling behaves as a semimetal phase. The introduction of spin-orbit coupling opens the bulk gap, transforming the system into a quantum spin Hall insulator~\cite{Bernevig2006}.
By depositing a lattice point into the square lattice structure, an additional flat band emerges away from the high-symmetry point $\Gamma$ in the bulk band, as shown in Fig.~\ref{fig1}(b).
The following sections explore various topological semimetal phases, including Dirac semimetal and Weyl semimetal, achieved through modulating the flat band.

To split flat bands and ensure Kramers degeneracy for arbitrary bands, we adopt bilayer models to construct Dirac semimetals. Interlayer coupling induces both upward and downward splitting of the bulk band while preserving time-reversal symmetry. As a result, we formulate the tight-binding Hamiltonian of the BHZ model in the bottom layer as follows:
\begin{align}
 H_{BHZ}^B &= (4t+ \epsilon - 2t \cos k_x - 2t \cos k_y ) \sigma_z \nonumber \\
 & + \lambda_x \sin k_x \sigma_x s_z - \lambda_y \sin k_y \sigma_y.
 \label{eq4}
\end{align}
The obtained Hamiltonian (\ref{eq4}) is derived by mirroring the top Hamiltonian along the y-axis. This setup provides the bilayer model Hamiltonian with real-space mirror symmetry $M_y$, which transforms $H(k_x,k_y)$ to $H(k_x,-k_y)$ by acting on the coordinate axes. Utilizing the tight-binding Hamiltonian in Eq.~(\ref{eq1}), we compute the band structures along the high symmetry points, illustrated in Fig.~\ref{fig1}(c). Here, we set the interlayer coupling to $\eta= 0.2$, and the other parameters are set to $t = 1, \epsilon=-1, \lambda_x=\lambda_y=1$, and $t_2=0.2$.
Due to time-reversal and inversion symmetry properties, the spin-up and spin-down bulk states remain degenerate and are indicated by the red solid and blue dashed lines, respectively.
Between the high symmetry points $X_1$ and $\Gamma$, a fourfold degenerate Dirac point is observed. This Dirac point is protected by both $\mathcal{T} \mathcal{I}$ symmetry and the real-space mirror symmetry $M_y$.
Specifically, $(\mathcal{T}  \mathcal{I} )^2 = -1$ ensures band Kramer degeneracy, while $M_y$ guarantees that the two Dirac points on the high symmetry line are symmetric about the $\Gamma$ point.
The location of two Dirac points in the Brillouin zone is marked by cyan dots as shown in Fig.~\ref{fig1}(d).
As a result, we establish the existence of two-dimensional Dirac semimetals that remain robust against spin-orbit coupling.
Notably, the Dirac point in this Dirac semimetal is located at the high symmetry line, unlike nonsymmorphic symmetry systems where it is located on the high symmetry points~\cite{Young2015}.
Furthermore, we calculate the variation of the Dirac point position with the coupling strength as shown in Figs.~\ref{fig1}(e1)-~\ref{fig1}(e4).
One can see that the Dirac point moves on the high symmetry line from $X_1$ to $\Gamma$ with increasing interlayer coupling strengths.
Such a move would change the distance between the two Dirac points in the Brillouin zone, thus extending the functionality of the Dirac points.

\begin{figure}
  \centering
  \includegraphics[width=8.6cm,angle=0]{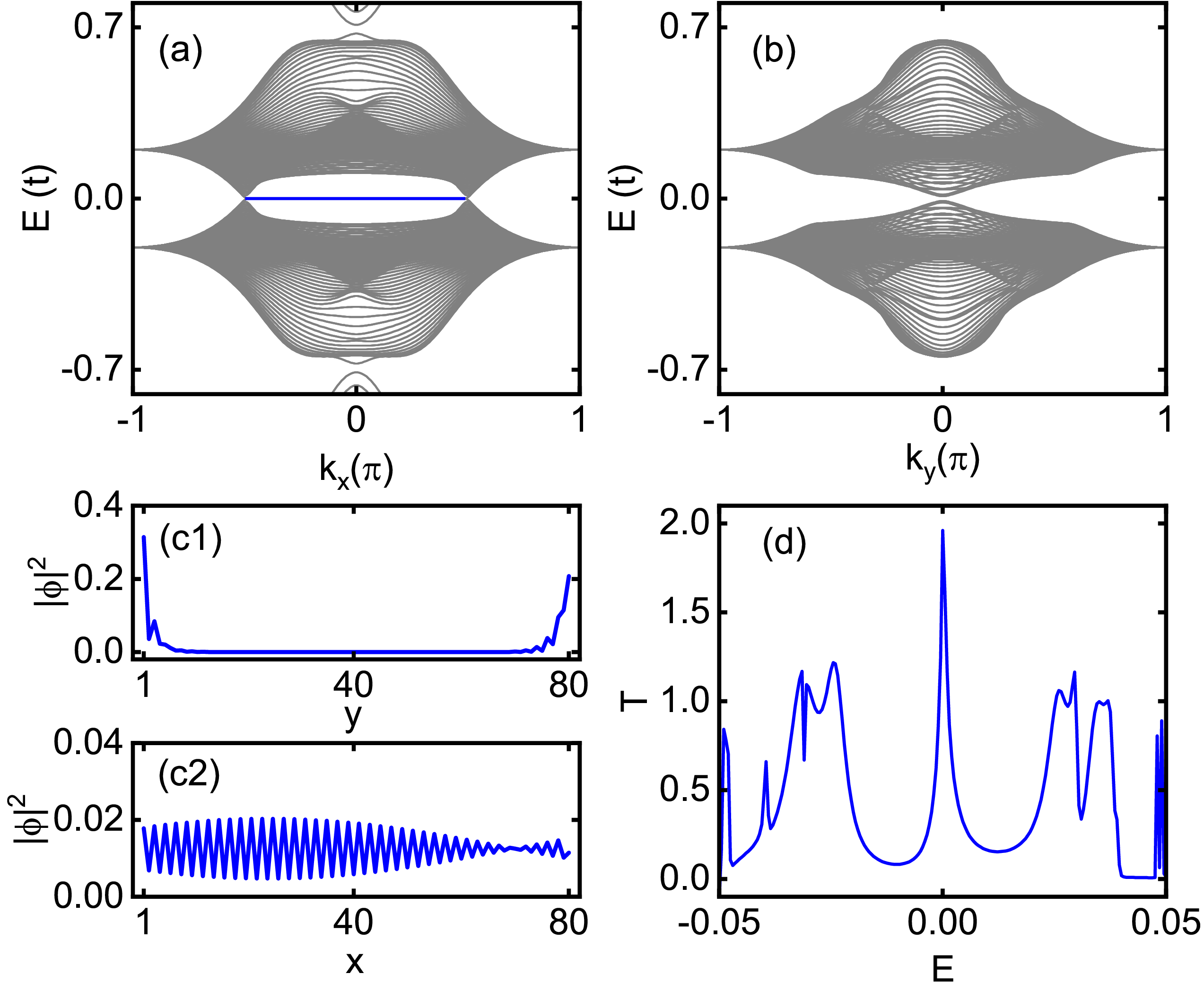}
  \caption{(a) Energy band structure of nanoribbons with periodic boundary conditions in the $x$-direction.
  The flat edge state of fourfold degenerate connects two Dirac points, highlighted with blue lines.
  (b) Energy band structure of nanoribbons with periodic boundary conditions in the $y$-direction.
  (c1) The wave function distributions of the four states near the zero energy for $k_x = 0$ and open boundaries in the $y$-direction.
  (c2) The wave function distributions of the four states near the zero energy for $k_y = 0$ and open boundaries in the $x$-direction.
  The parameters are set to $t = 1, \epsilon=-1, \lambda_x=\lambda_y=1, t_2=0.2$, and $\eta=0.2$.
  (d) Tunneling coefficient for incident electrons versus incident energy E with coupling strength $t_1=0.15$.
  }
  \label{fig2}
\end{figure}

In two-dimensional materials, beyond the intriguing properties of Dirac semimetals, nontrivial edge states can exhibit unique transport characteristics. To investigate this boundary scenario, we analyze the energy band structure of the Dirac semimetal nanoribbon, as depicted in Fig.~\ref{fig2}. Figure~\ref{fig2}(a) illustrates the energy band structure of the nanoribbon with an open boundary in the y-direction, employing parameters $t = 1, \epsilon=-1, \lambda_x=\lambda_y=1, t_2=0.2, \eta=0.2$, and a nanoribbon width of $N = 80$. Notably, one can observe a fourfold degenerate flat edge state in the $k_x$ direction highlighted by the blue line, connected by the projections of the two Dirac points.
Considering the position of the Dirac point in the Brillouin zone [see Fig.~\ref{fig1}(d)], in Fig.~\ref{fig2}(b) we plot the energy band structure dependent on $k_y$, with the $x$ direction set as the open boundary. Remarkably, there are no edge states along the $k_y$ direction. This absence arises because that the two Dirac points are separated in the $k_x$ direction, but they are mixed along the $k_y$ direction.

To show the edge states more clearly, we plot the wave function distributions of the four states near the zero energy for $k_x = 0$ and open boundaries in the $y$-direction, as shown in Fig.~\ref{fig2}(c1). The wave function is mainly localized at two ends, proving the existence of edge states in the $k_x$ direction. As a comparison, in Fig.~\ref{fig2}(c2), we show the wave function distributions of the four states near the zero energy at $k_y = 0$ with open boundaries in the $x$ direction. This is the bulk state and not the edge state. Consequently, in this setup, the Dirac semimetal resulting from interlayer coupling exhibits nontrivial edge states only in the $k_x$ direction. This orientation-dependent edge state holds significant potential for applications in anisotropic electronic devices. Moreover, interlayer coupling can alter the distance from the Dirac point to the $\Gamma$ point, thereby influencing the length of the edge states in momentum space.

Although the flat edge states are distributed at the edges [see Fig.~\ref{fig2}(c1)], they manifest as bound states. To verify this, we calculate the transport properties of a two-terminal system by connecting the left and right sides of a semimetal system with bilayer quantum anomalous Hall insulator (BQAHI) leads. In this setup, two incident electrons with energy $E$ flow from the edge of the left terminal to the BQAHI-semimetal-BQAHI junction. The Hamiltonian of the left (right) lead $H_{L(R)}$ can be similarly described by Eqs.~(\ref{eq1})-(\ref{eq3}), where $H^B$ is replaced by $H^T$ in Eq.~(\ref{eq1}) and the extra term $B_{z}\sigma_{z}s_{z}+\mu_1 \sigma_{0}s_{0}$ is added to Eq.~(\ref{eq3}). $B_{z}$ and $\mu_1$ represent the magnitude of the out-of-plane magnetic field and chemical potential, respectively. The parameters of the leads are set as $t = 1, \epsilon=-1, \lambda_x=\lambda_y=1$, $t_2=0$, $\eta=0$, $B_{z}=1.8$, and $\mu = \mu_1= 0.05$. Employing the nonequilibrium Green's function~\cite{Fisher1981, Metalidis2005, Sun2009}, we can obtain the tunneling coefficient $T(E)={\rm Tr}[{\rm \Gamma}_{L} G^{r}_{} {\rm \Gamma}_{R} (G^{r}_{})^{\dagger}]$, where ${\rm \Gamma}_{L(R)}(E)= i[\Sigma_{L(R)}^{r}-(\Sigma_{L(R)}^{r})^{\dagger}]$ is the linewidth function of the left (right) lead. $G^{r}(E)=[E-H_{C}-\Sigma_{L}^{r}-\Sigma_{R}^{r}]^{-1}$ is the retarded Green's function, where $H_{C}$ is the Hamiltonian of the central scattering region. The retarded self-energy terms contributed by the left and right leads are $\Sigma_{L}^{r}=H_{CL}g^{r}_{L}H_{LC}$ and $\Sigma_{R}^{r}=H_{RL}g^{r}_{R}H_{RC}$, where $g^{r}_{L}$ and $g^{r}_{R}$ are the surface Green's function of the semi-infinite leads~\cite{Lee1981}. $H_{CL}=H_{LC}^{\dagger}$ ($H_{CR}=H_{RC}^{\dagger}$) is the coupling Hamiltonian between the central scattering region and the left (right) lead, which is equal to coupling strength $t_{1}$ times the hopping part of the leads' Hamiltonian. In Fig.~\ref{fig2}(d), we plot the tunneling coefficient as a function of incident energy $E$ with the coupling strength $t_1=0.15$. It can be seen from Fig.~\ref{fig2}(d) that a resonance peak $T=2$ appears near $E=0$, which proves that the flat edge states are indeed bound.

\begin{figure}
  \centering
  \includegraphics[width=8.6cm,angle=0]{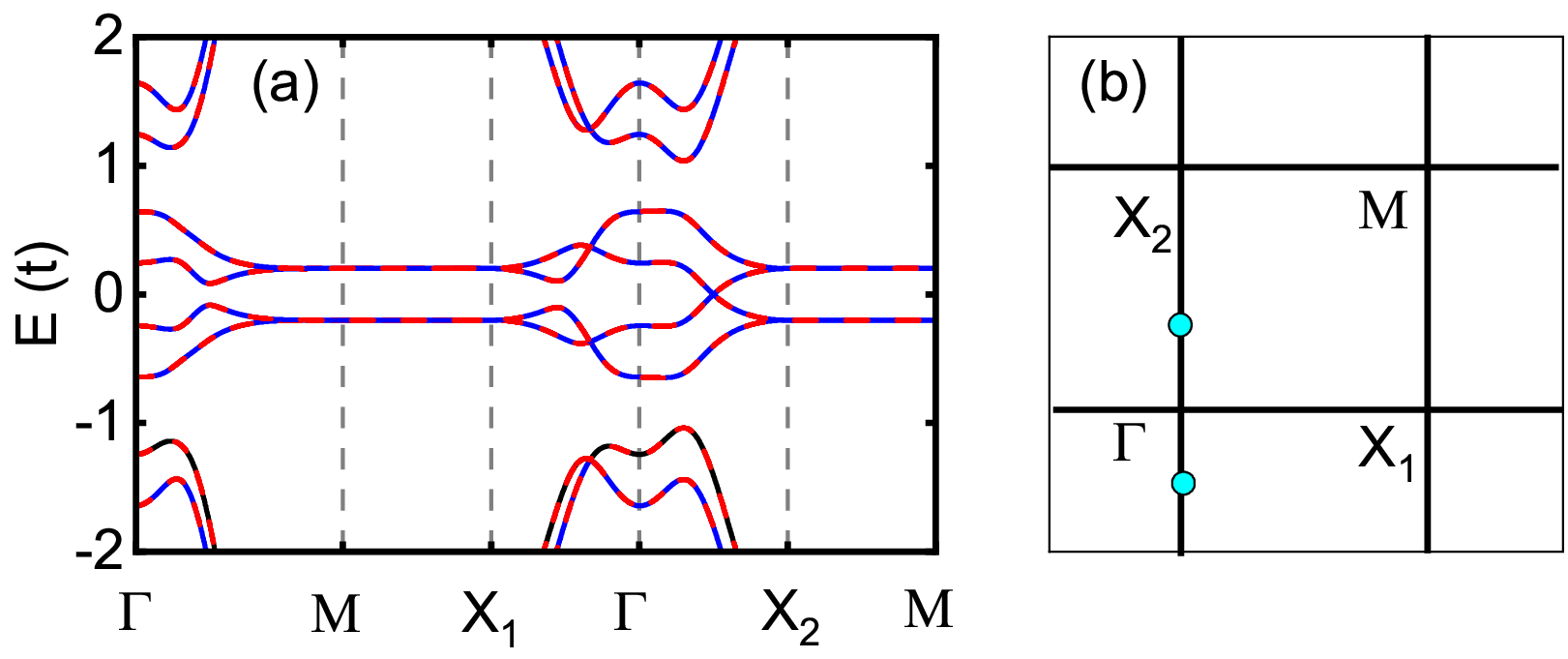}
  \caption{ (a) The bulk energy bands structures for bilayer modified BHZ model along the high-symmetry points with Dirac points protected by $\mathcal{T} \mathcal{I}$ symmetry and mirror symmetry $M_x$.
  Red solid lines indicate the spin-up band and the blue dashed lines indicate the spin-down band. 
  (b) Dirac points in the Brillouin Zone are marked in cyan.
  }
  \label{fig3}
\end{figure}

Since the edge state connects the projections of the two Dirac points, the position of the Dirac points in the two-dimensional system affects its unique transport properties.
We now vary the position of the Dirac points by using different bottom Hamiltonians $H^B$ to satisfy different mirror symmetries.
Above, we discussed the bilayer model that satisfies the mirror symmetry $M_y$. Naturally, we reset the bilayer model protected by the mirror symmetry $M_x$.
Therefore, the Hamiltonian of the bottom BHZ model is written as:
\begin{align}
 H_{BHZ}^B&=(4t + \epsilon - 2t \cos k_x - 2t \cos k_y ) \sigma_z \nonumber \\
 & - \lambda_x \sin k_x \sigma_x s_z + \lambda_y \sin k_y \sigma_y.
 \label{eq5}
\end{align}
The parameters are the same as in Fig.~\ref{fig2}.
The bulk energy band structure of the bilayer system is depicted in Fig.~\ref{fig3}(a).
One can see that the bulk bands intersect in a linear dispersion at the Fermi energy between the high symmetry points $\Gamma$ and $X_2$.
Due to time-reversal and inversion symmetries, the bands exhibit Kramers degeneracy for all $k$ values.
Consequently, the point of energy band closure is fourfold degenerate, indicative of a Dirac semimetal.
Notably, the two Dirac points in the Brillouin zone reside in the same $k_x$ space, marked by cyan dots in Fig.~\ref{fig3}(b).
This configuration results in the longest edge state in the $k_y$ periodic boundary, which smoothly contracts with the boundary direction and completely disappears in the $k_x$ periodic boundary.
In the $k_x$ periodic boundary, the two Dirac points project onto each other, leading to a zero length for the edge state, effectively causing it to vanish.

\begin{figure}
  \centering
  \includegraphics[width=8.6cm,angle=0]{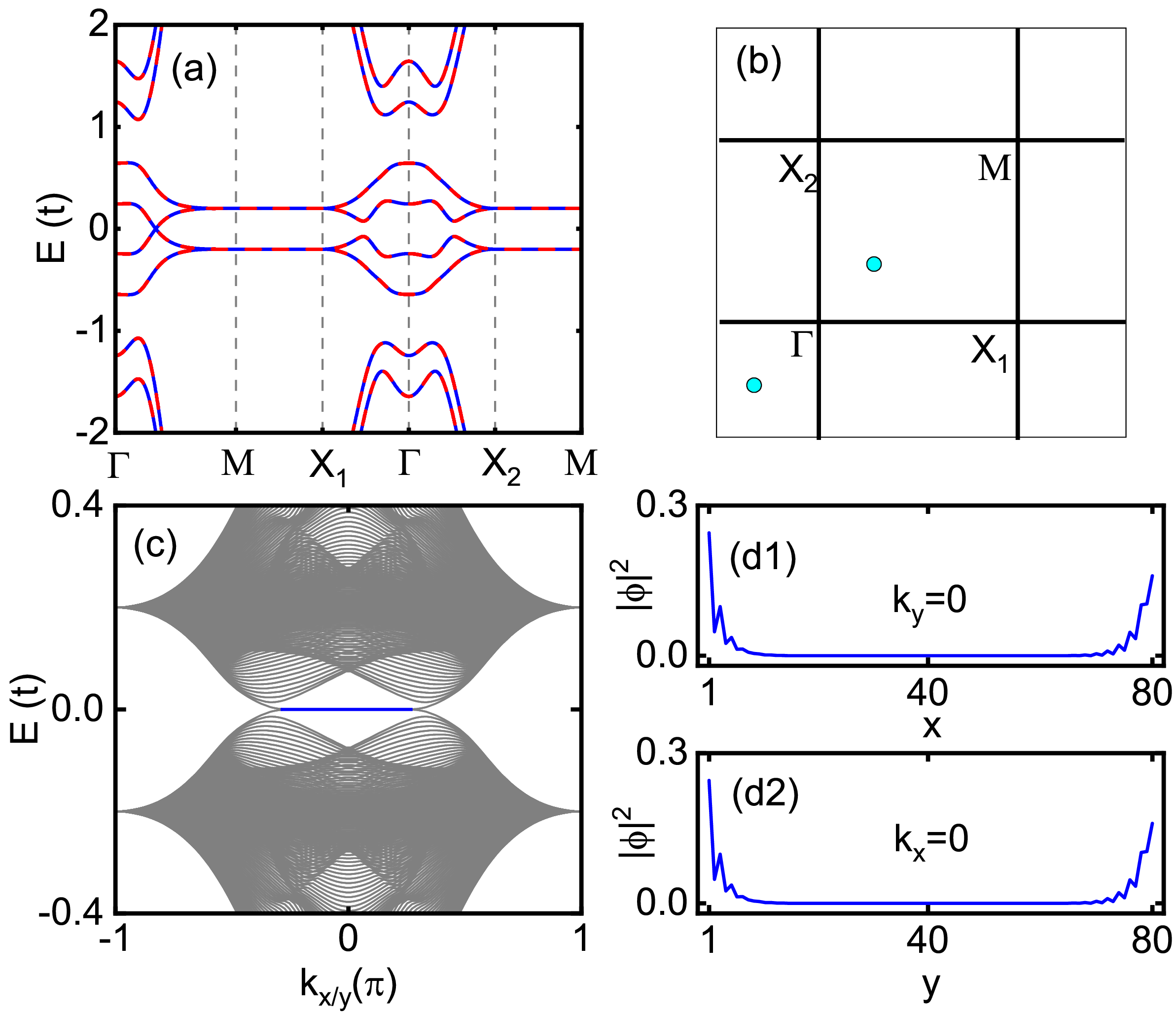}
  \caption{(a) The bulk energy bands structures for bilayer modified BHZ model along the high-symmetry points with Dirac points protected by $\mathcal{T} \mathcal{I}$ symmetry.
  Red solid lines indicate the spin-up band and the blue dashed lines indicate the spin-down band.
  (b) Dirac points in the Brillouin Zone are marked in cyan.
  (c) Energy band structure of nanoribbons with periodic boundary conditions in the $x$ (or $y)$-direction.
  (d1) The wave function distributions of the four states near the zero energy for $k_y = 0$ and open boundaries in the $x$-direction.
  (d2) The wave function distributions of the four states near the zero energy for $k_x = 0$ and open boundaries in the $y$-direction.
  }
  \label{fig4}
\end{figure}
To delve deeper into edge states in Dirac semimetals, we introduce the mirror symmetry $M_{xy}$ to ensure that the two Dirac points remain distinct in both the $k_x$ and $k_y$ directions.
Therefore, we define the bottom BHZ model as follows:
\begin{align}
 H_{BHZ}^B&=(4t + \epsilon - 2t \cos k_x - 2t \cos k_y) \sigma_z \nonumber \\
 & + \lambda_x \sin k_y \sigma_x s_z + \lambda_y \sin k_x \sigma_y.
 \label{eq6}
\end{align}
In this scenario, the Hamiltonian of the bilayer system adheres to the real-space mirror symmetry $M_{xy}$.
As depicted in Fig.~\ref{fig4}(a), a fourfold degenerate bulk band linear closure between $\Gamma$ and $M$ arises due to the combined effects of $\mathcal{T} \mathcal{I}$ symmetry and mirror symmetry $M_{xy}$.
In Fig.~\ref{fig4}(b), the two Dirac points symmetric about the $\Gamma$ point in the Brillouin zone are marked in cyan.
These Dirac points do not mix in both the $k_x$ and $k_y$ directions.

To verify whether the edge states connect the mirror symmetry $M_{xy}$ protected Dirac points, we calculate the energy band structure along the $k_y$ direction for the $x$-direction open boundary nanoribbons, as shown in Fig.~\ref{fig4}(c).
The parameters are the same as in Fig.~\ref{fig2}.
We observe that the fourfold degenerate flat band connects the projections of the two Dirac points.
The wave function distribution of the zero energy state corresponding to the flat band at $k_y = 0$ is shown in Fig.~\ref{fig4}(d1).
It is mainly localized at two ends of the $x$-direction, which verifies that the edge state connects the projections of the two Dirac points in the $k_y$-direction.
The energy band structure with the $k_x$ direction, when the $y$ direction is an open boundary, is the same as in Fig.~\ref{fig4}(c).
The wave function distribution in Fig.~\ref{fig4}(d2) demonstrates that the edge states along the $k_x$ direction connect the projections of the Dirac points.
Consequently, the edge states connecting the two Dirac points can manifest in either the $k_x$ or $k_y$ direction.

The above three cases illustrate different arrangements of the Dirac points:
(i) In the first case, the two Dirac points are mixed in the $k_y$ direction, protected by the symmetry $M_y$, resulting in edge states existing exclusively in the $k_x$ direction.
(ii) In the second case, the two Dirac points are mixed in the $k_x$ direction, protected by the symmetry $M_x$, leading to edge states existing solely in the $k_y$ direction.
(iii) Lastly, in the third case, the two Dirac points do not mix in either the $k_x$ or $k_y$ directions, protected by the symmetry $M_{xy}$, allowing edge states to exist in both the $k_x$ and $k_y$ directions.
The coexistence of Dirac points and tunable edge states at the Fermi energy may lead to peculiar electron transport properties.

\begin{figure}
  \centering
  \includegraphics[width=8.6cm,angle=0]{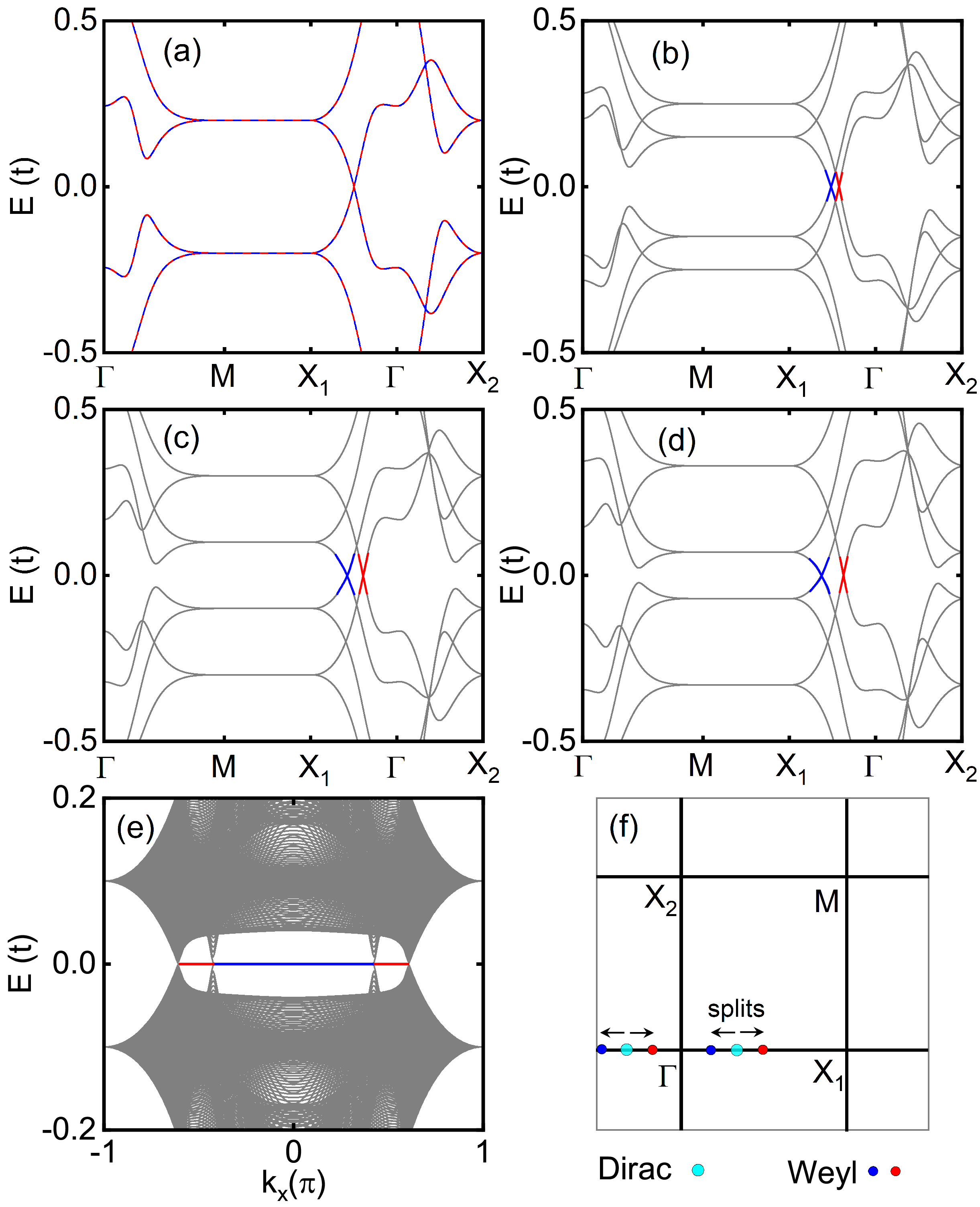}
  \caption{ 
  (a) The enlarged view of the Fermi energy neighborhood in Fig.~\ref{fig1}(c).
  (a-d) The bulk energy band structure along the high symmetry points of the bilayer modified BHZ model with magnetic deposition of different strengths,
  for (a) $B_x=0$, (b) $B_x=0.05$, (c) $B_x=0.1$, and (d) $B_x=0.13$.
  Blue and red bands crossed indicate Weyl points.
  (e) Energy band structure of nanoribbons with open boundary conditions in the y-direction.
  The blue (red) flattened band is the edge state of the fourfold (twofold) degenerate.
  (f) The Dirac points (cyan dots) in the Brillouin zone splits into two Weyl points (blue and red dots) due to magnetic deposition.}
  \label{fig5}
\end{figure}
\section{two-dimensional Weyl Semimetals in bilayer systems}\label{SecIV}

It is well known that Dirac points in semimetals are protected by time-reversal symmetry and inversion symmetry. Fundamentally, the prevailing mechanism for Weyl point formation is the broken time-reversal or inversion symmetry of the Dirac semimetal~\cite{Xu2011}. To show that the Dirac semimetal of the bilayer modified BHZ model can be designed as a time-reversal symmetry-broken Weyl semimetal, we introduce magnetic deposition sites, i.e., $B_{x} \neq 0$.

To demonstrate the Dirac point splitting due to magnetic deposition, in Figs.~\ref{fig5}(a)-\ref{fig5}(d) we plot the bulk band structure along the high symmetry points at different magnetic strengths, respectively.
The magnetization strengths are set to $B_x=0, B_x=0.05, B_x=0.1,$ and $B_x=0.13$, respectively, and the other parameters are the same as in Fig.~\ref{fig1}(c).
Fig.~\ref{fig5}(a) is an enlarged view of Fig.~\ref{fig1}(c) in the vicinity of the Fermi energy, placed here to show the Dirac point splitting process more clearly.
The spin-up and spin-down bulk states remain degenerate and are indicated by the red solid and blue dashed lines, respectively.
From the comparisons in Figs.~\ref{fig5}(a)-\ref{fig5}(d), one can see that the fourfold degenerate Dirac point splits into two Weyl points (blue and red bands crossing) and that the degree of splitting increases with increasing magnetization intensity.

Now, we turn our attention to the exploration of nanoribbon edge states.
In Fig.~\ref{fig5}(e), we plot the energy band structure of nanoribbon with open boundaries in the y-direction, with the parameters being chosen as $t = 1, \epsilon=-1, \lambda_x=\lambda_y=1, t_2=0.2, \eta=0.2$, and $B_x=0.1$.
We observe that the fourfold degenerate flat edge states (blue line) do not vanish, and that the two Weyl points split by a Dirac point are connected by the twofold degenerate flat edge states (red lines).
The above results demonstrate that each Dirac point splits into two Wely points.
In Fig.~\ref{fig5}(f), a schematic of a two-dimensional Brillouin zone is plotted.
Cyan dots on the high symmetry path indicate Dirac points in the bilayer system without magnetism.
Once magnetism is introduced each Dirac point splits along a high symmetric path into two Weyl points, denoted by blue and red dots, respectively.
Notably, the deposition in any magnetic direction induces the bilayer system to be a two-dimensional Weyl semimetal.
Furthermore, we also find that the introduction of magnetic deposition into the monolayer modified BHZ model will lead to Weyl semimetals with edge states, as discussed in detail in Appendix.

\section{DISCUSSIONS AND CONCLUSIONS}\label{SecV}
We employ the bilayer modified BHZ model to establish a diverse array of Dirac semimetal phases with edge states.
The modified model is achieved by introducing deposition into vacancies within the BHZ model, resulting in a two-site two-dimensional system featuring flat-like bands within the bulk bands.
By coupling two modified BHZ models with distinct configurations,
we create a bilayer model capable of satisfying mirror symmetries $M_x$, $M_y$, or $M_{xy}$.
These symmetries can fix the direction of the Dirac points and the interlayer coupling strength determines the distance between the two Dirac points.
Furthermore, the edge states in Dirac semimetals serve to connect two Dirac points within the Brillouin zone.
Quantized resonance peaks prove the existence of edge states.
The symmetry $M_y$ ($M_x$) ensures the existence of edge states in the $k_x$ ($k_y$) direction, whereas $M_{xy}$ guarantees edge states in both the $k_x$ and $k_y$ directions, enabling the design of electronic devices with anisotropy at each edge.
Furthermore, we find that magnetic deposition leads to cleavage of each Dirac point into two Weyl points.
The monolayer modified BHZ model can be obtained by uniform depositing non-magnetic atoms such as nitrogen or boron on the HgTe/CdTe interstitial sites~\cite{George2010, Johnson2014, Kunene2022, Park2024}.
Coupling the two modified BHZ models one can observe Dirac semimetals with edge states.
Finally, we find that magnetic deposition induces the monolayer system to transition into a two-dimensional Weyl semimetal with edge states.

\section*{acknowledgments}
This work was financially supported by the National Natural Science Foundation of China (Grants No. 12074097, No. 12374034, and No. 11921005),
Natural Science Foundation of Hebei Province (Grant No. A2024205025),
the Innovation Program for Quantum Science and Technology (Grant No. 2021ZD0302403),
and the Strategic Priority Research Program of the Chinese Academy of Sciences (Grant No. XDB28000000).


\section*{APPENDIX: Weyl semimetal in magnetically deposited monolayer system} \label{APPENDIX}

\begin{figure}
  \centering
  \includegraphics[width=8.6cm,angle=0]{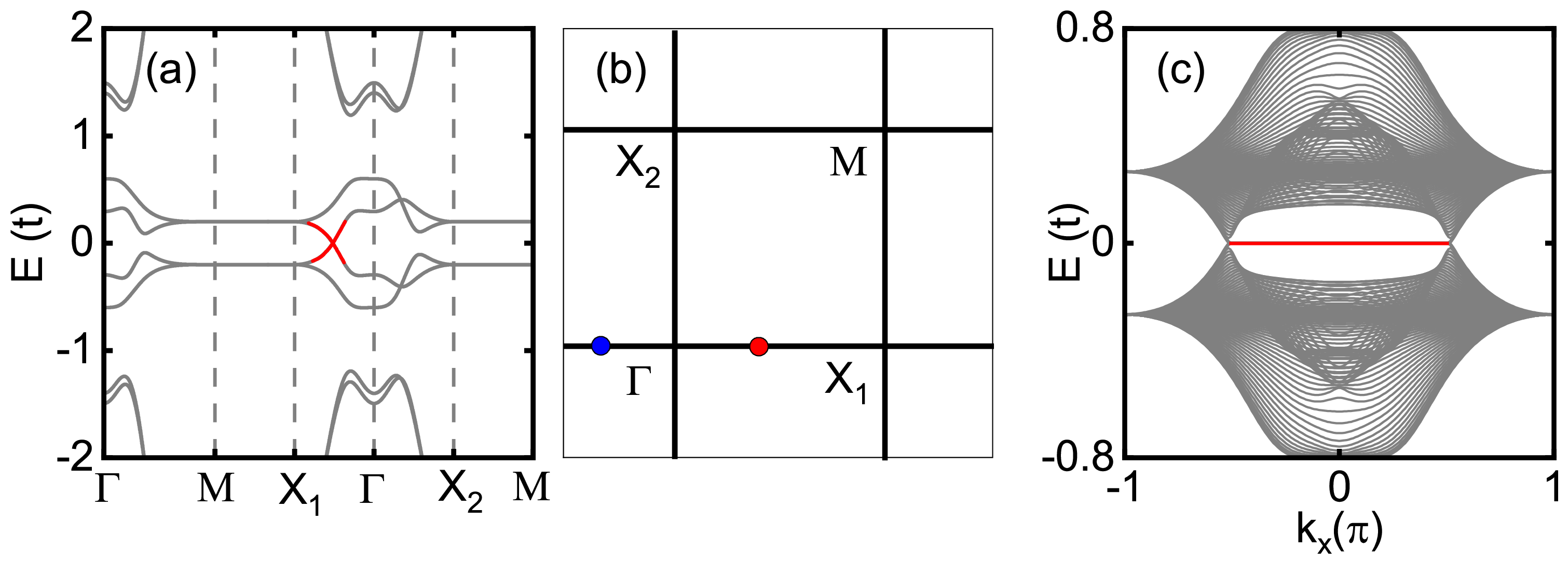}
  \caption{ (a) Energy band structure of a single-layer modified BHZ model with magnetic deposition.
  (b) Weyl points in the Brillouin Zone are marked in blue and red.
  (c) Energy band structure of nanoribbons with periodic boundary conditions in the $x$-direction.
  The flat edge state of twofold degenerate connects two Weyl points, highlighted with red line.
  }
  \label{fig6}
\end{figure}

In our previous study~\cite{Liu2024}, the effective Hamiltonian equations demonstrated equivalence between the in-plane magnetism acting on the spin and the coupling between the layers.
However, the magnetic field differs in that it breaks time-reversal symmetry, thereby disrupting Kramer's degeneracy.
As a result, the crossing of the energy bands at the Fermi energy must be twofold degenerate, forming a Weyl-type semimetal.
We now investigate the effect of magnetic deposition ($B_x=0.2$) on two-dimensional monolayer systems.
In Fig.~\ref{fig6}(a), we compute the bulk band structure using Hamiltonian (\ref{eq2}) with parameters $t = 1, \epsilon=-1, \lambda_x=\lambda_y=1, t_2=0.2$.
The twofold degenerate band crossover is observed between the high symmetry points $\Gamma$ and $X_1$.
The blue and red dots in Fig.~\ref{fig6}(b) indicate the location of the Weyl point in the Brillouin Zone.

It's evident that the two Weyl points are mixed in the $k_y$ direction but not in the $k_x$ direction.
Consequently, the edge state connecting the two Weyl points exists exclusively in the nanoribbon with periodic boundary conditions in the $k_x$ direction.
Fig.~\ref{fig6}(c) illustrates the energy band structure of the nanoribbon along the $k_x$ direction, showcasing the flattened edge states connecting the projections of the two Weyl points.
Moreover, we demonstrate that an in-plane magnetic field in any direction can induce monolayer samples into Weyl semimetals with edge states.

\end{document}